\documentclass[11pt]{article}

\usepackage{amsmath,amssymb}
\usepackage{graphicx}
\usepackage{epstopdf}
\usepackage{epsfig}

\input epsf

\newcommand{\be}{\begin{equation}}
\newcommand{\ee}{\end{equation}}
\newcommand{\bea}{\begin{eqnarray}}
\newcommand{\eea}{\end{eqnarray}}
\newcommand{\ba}{\begin{array}}
\newcommand{\ea}{\end{array}}

\newcommand{\eq}[1]{(\ref{#1})}

\renewcommand{\title}[1]{\vbox{\center\LARGE{#1}}\vspace{5mm}}
\renewcommand{\author}[1]{\vbox{\center#1}\vspace{5mm}}

\numberwithin{equation}{section}

\begin{document}

\begin{titlepage}
\begin{flushright}
\end{flushright}

\vfill

\begin{center}
{\large \bf Breathing Vacuum Bubbles in Five-Dimensional
Gauss-Bonnet Gravity} \vfill { Feng-Li Lin$^a$\footnote{\tt
linfengli@phy.ntnu.edu.tw}, } {Chien-Hsun Wang$^a$\footnote{\tt
696410112@phy.ntnu.edu.tw}, }
 and
{ Chen-Pin Yeh$^b$\footnote{\tt chenpinyeh@gmail.com} }

\bigskip
{\it $^a$Department of Physics, National Taiwan Normal University,
Taipei, 116, Taiwan\\
}

and

{\it $^b$Department of Physics, National Taiwan University,
Taipei, 106, Taiwan\\
}

\end{center}

\vfill
\begin{abstract}
We study the dynamics of bubble wall in five-dimensional
Gauss-Bonnet gravity in the thin-wall approximation. The motion of the domain wall is determined
by the generalized Israel junction condition. We find a solution
where the wall of true vacuum bubble is oscillating, the breathing bubbles. We briefly
comment on how this breathing vacuum bubble can affect the
analysis of the string theory landscape.
\end{abstract}
\vfill

\end{titlepage}

\setcounter{footnote}{0}

\section{Introduction}
In the theory with multiple meta-stable vacua, string theory for
example, our universe is very likely to tunnel from other vacua
with higher cosmological constants. The process is similar to the
first order phase transition, where the bubble with different
vacuum energy will nucleate from the original vacuum. When couple
to the gravity, the vacuum with positive energy will expand
exponentially. This leads to an interesting cosmology of eternal
inflation \cite{eternal inflation}.

In Einstein gravity, the geometry and dynamics of the bubble
nucleation have been studied extensively, see \cite{Blau:1986cw}
and references therein. The motion of the bubble wall, in the thin
wall approximation is determined by the Israel junction condition
\cite{Israel}. There are various kinds of solutions, but they fall
into two main categories. When the wall tension can't compensate the
energy occupied by the bubble's volume, it will eventually shrink
to zero size. Otherwise the bubble wall will eventually expand and
quickly approach the speed of light.

However, these are not the most general solutions. The bubble walls
can oscillate and neither shrink to zero size nor asymptote to the
light cone. This kind of breathing bubble solution, in Einstein
gravity has been found in the paper \cite{Guen:2008}. The solution
in their case requires the bubble wall to have non-standard
equation of state. Similar solution is also found in the study of
brane-world cosmology, where our universe lives on the thin brane
embedded in the higher dimension space-time. The Israel junction
condition here determined the scale factor of the brane-world. In
\cite{Kraus:1999}, the author found that, in order to get the
breathing type solution the brane must be embedded in the Anti-de
Sitter space.

The Gauss-Bonnet gravity is a particular higher derivative
correction to Einstein gravity, which leaves the equation of
motion remain second order. It is topological in four dimension
and will not affect the dynamics. In this paper we instead study
the bubble dynamics in the five-dimensional(5D) Gauss-Bonnet
gravity. The energy momentum tensor of the bubble wall is assumed
to be the form of ordinary scalar field. In many cases the
solution is qualitatively the same as in Einstein gravity. However
in a wide range of parameter space, we find the breathing bubble
solution for the true vacuum bubble with positive vacuum energy.
At first sight, it may seem strange to consider the bubble in
five-dimensional Gauss-Bonnet gravity. However in the UV complete
theory of gravity. It is generic to have higher derivative
corrections to Einstein gravity. When we consider the evolution of
the early universe, the effect of higher derivative terms can play
an important role. Moreover, landscape of vacua in string theory
\cite{Landscape} also contains those with effectively space-time
dimension higher than four. The transition between bubbles with
different dimension is also possible \cite{Vilenkin:2009}. There
is a possibility that our universe tunnelled from such higher
dimensional bubbles. Thus it may be too hasty to restrict ourself
on 4D Einstein gravity. For example, it needs to be more careful
when discussing the measure problem, the problem to assign the
probability to different vacuum bubbles.(See \cite{Vilenkin:2006}
for review). As it usually assumes that the bubble wall approaches
the light cone.

In section two, we will first review the bubble wall dynamics in
5D Einstein gravity. Then we turn to the 5D Gauss-Bonnet gravity
and find the breathing bubbles.  We conclude in section three and
point out the possible effect of breathing bubble solution on the
measure problem. In Appendix A, we discuss the difficulty of the
thin wall approximation in general higher derivative gravity other
than Gauss-Bonnet gravity. However for showing the effects of
higher derivative corrections, we believe that the Gauss-Bonnet
gravity is sufficient.

\section{Bubble dynamics in 5D Gauss-Bonnet gravity}
In this section, we first study the bubble dynamics in 5D Einstein
gravity. The solutions will not be much different from 4D Einstein
gravity if we assume the spherical symmetry. Nevertheless the
results can then be used to compare with the ones from the 5D
Gauss-Bonnet gravity. We then turn to bubble dynamics in 5D
Gauss-Bonnet gravity. One complication is that the analogy of
particle dynamics in 1D effective potential is not obvious in the
Gauss-Bonnet counterpart. Here we use the phase space analysis to
study the bubble wall trajectories.

\subsection{Vacuum bubble dynamics in 5D Einstein gravity}

We start with the 5D Einstein-Hilbert action

\be
S=\frac{1}{2\kappa^{2}}\int d^{5}x \sqrt{-g}\; [R-2\Lambda]
\ee
where $\kappa^2$ is the 5D Newton constant, and $\Lambda$ is
the cosmological constant. Assuming spherical symmetry for both
the bulk space-time and the bubble wall, we can solve the Einstein
equation and get the bulk metric solution in the following form

\be
\label{metricb}
ds^{2}=-f_{in,out}(r)dt^{2}+f_{in,out}(r)^{-1}dr^{2}+r^{2}d\Omega^{2}_{3}
\ee
where $f_{in,out}(r)$'s are the harmonic function for the
inside and outside of the bubble.

Given the inside and outside forms of the metric, the bubble wall
dynamics is dictated by the Israel junction condition
\cite{Israel}, which can be derived from Einstein equation and
takes the form as

\be
[K_{ab}-Kh_{ab}]^{+}_{-}=-\kappa^{2} S_{ab}
\ee
where $K_{ab}$
is the extrinsic curvature of the bubble wall, and $S_{ab}$ is the
stress tensor of matters on the bubble wall. According to
spherical symmetry, we can choose the induced metric on the bubble
wall in the following form

\be
\label{metricw} ds^{2}=-d\tau^{2}+r^{2}(\tau)d\Omega^{2}_{3}.
\ee
Then, the stress tensor of the wall takes the form

\be
\label{stressw}
S_{ab}=-\sigma h_{ab}
\ee
where $\sigma$ can be
understood as the tension of the wall.

Choosing the gauge $t=\tau$, and plugging the wall metric into the
Israel junction condition to get the dynamical equation for the
bubble radius $r(t)$. Due to the spherical symmetry, the only
independent equation (the $tt$ component only) takes the form
\be
\label{Einstein1}
-\frac{\kappa^{2}\sigma}{3}r=\sqrt{\dot{r}^{2}+f_{out}}-\sqrt{\dot{r}^{2}+f_{in}}\;,
\ee
where $\dot{r}:={dr \over dt}$. This takes the same form as
for the 4D Einstein gravity.

\begin{figure} \epsfxsize=200pt
\centerline{\epsfbox{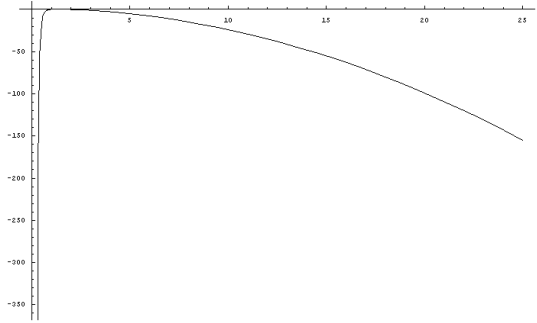}} \caption{Effective potential for true
vacuum bubble in Einstein gravity. } \label{fig:1}
\end{figure}

To be more specific, we now consider the dynamics of the true
vacuum bubble, that is
\be
f_{in}=1\;, \qquad f_{out}=1-{\Lambda
\over 6}r^2-{\mu \over r^2}\;,
\ee
which describe an Minkowski
bubble surrounded by the de Sitter space, and $\mu$ is the mass of
the bubble measured by an asymptotic outside observer. After some
rearrangement, \eq{Einstein1} can be put into form of particle
dynamics
\be
\label{particle1}
\frac{1}{2}\dot{r}^{2}+V_{eff}=0
\ee
with an effective potential as
\be
\label{particle2}
V_{eff}=\frac{1}{2}-{9\over 8}\frac{\left(({\kappa^4\sigma^2 \over
9}-{\Lambda \over 6})r^{2}+\mu/r^{2}\right)^{2}}{\kappa^4 \sigma^2
r^{2}}
\ee
The form of the effective potential is shown in Figure
\ref{fig:1}. This is quite similar to case for 4D Einstein
gravity.

By inspecting the form of the effective potential, it is easy to
see there are three kinds of solutions of \eq{particle1} and
\eq{particle2} characterized by the behaviors of $r(t)$. They are
bounded, bounce and monotonic ones, which correspond to the
shrinking, bouncing and ever-expanding bubbles, respectively.
Again, these solutions are similar to the ones in 4D Einstein
gravity.

\begin{figure}
\epsfxsize=250pt \centerline{\epsfbox{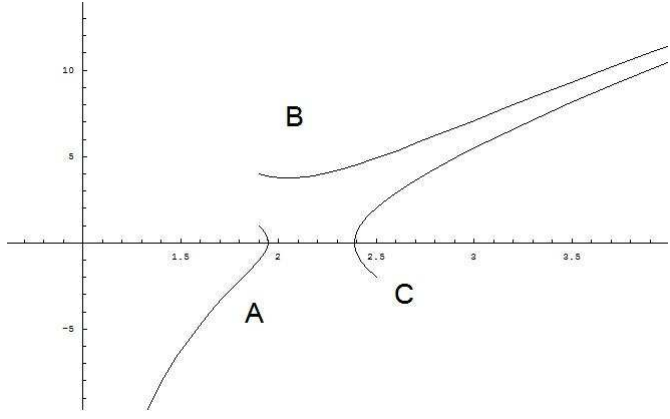}} \caption{True
vacuum bubble solutions in phase space for 5D Einstein gravity. A:
bounded solution. B: monotonic solution. C: bounce solution.
}\label{fig:2}
\end{figure}

One can also change the particle dynamics to the Hamilton
formalism by regarding \eq{particle1} as a Hamiltonian constraint,
i.e., $H=0$. By tuning the initial conditions, one can then
numerically characterize the above 3 different solutions in the
phase space $(q=r,p=\dot{r})$ as shown in Figure \ref{fig:2}.It
seems redundant to numerically solve the bubble wall dynamics in
the Hamilton's formalism. However, it turns out to be quite
essential for the case of Gauss-Bonnet gravity, in which particle
dynamics is no longer quadratic and is hard to visualize the
solution by just inspecting the effective potential.

\subsection{Junction condition of 5D Gauss-Bonnet gravity}
We now move to the bubble dynamics in the 5D Gauss-Bonnet gravity
by starting with the action

\be S=\frac{1}{2\kappa^{2}}\int d^{5}x
\sqrt{-g}[R-2\Lambda+\alpha {\cal L}_{GB}]
\ee
where $\alpha$ is
the Gauss-Bonnet coupling and the form of Gauss-Bonnet term is

\be
{\cal
L}_{GB}=R^{2}-4R_{\mu\nu}R^{\mu\nu}+R_{\mu\nu\rho\sigma}R^{\mu\nu\rho\sigma}\;.
\ee
Due to the Gauss-Bonnet term, the junction condition for the
bubble wall becomes more involved as following \cite{Davis:2003},
and it takes the form as

\be
[K_{ab}-Kh_{ab}]^{+}_{-}+2\alpha[3J_{ab}-Jh_{ab}+2P_{acdb}K^{cd}]^{+}_{-}=-\kappa^{2}
S_{ab}
\ee
where

\be
J_{ab}=\frac{1}{3}(2KK_{ac}K^{c}_{b}+K_{cd}K^{cd}K_{ab}-2K_{ac}K^{cd}K_{db}-K^{2}K_{ab})
\ee
and

\be
P_{abcd}=R_{abcd}+2R_{b[c}g_{d]a}-2R_{a[c}g_{b]b}+Rg_{a[c}g_{d]b}\;.
\ee

Again assuming spherical symmetry for the bubble wall with the
bulk and wall metrics \eq{metricb} and \eq{metricw}, respectively;
then $tt$ component of the junction condition in the $t=\tau$
gauge can be reduced to

\be \label{EOM}
-\kappa^{2}\sigma=\Bigl[3\frac{\sqrt{\dot{r}^{2}+f(r)}}{r}
-2\alpha\Bigl(2\frac{\Bigr(\sqrt{\dot{r}^{2}+f(r)}\Bigl)^{3}}{r^{3}}-6(1+\dot{r}^{2})
\frac{\sqrt{\dot{r}^{2}+f(r)}}{r^{3}}\Bigr)\Bigr]^{+}_{-}\nonumber
\ee where we have used \eq{stressw} for the stress tensor of the
matter on the wall.  It is clear that the "particle dynamics" for
$r(t)$ based on \eq{EOM} is no longer quadratic, and in general it
can be transformed into the following form
\begin{eqnarray}
X(r)\dot{r}^{2}+Y(r)\dot{r}^{4}+Z(r)\dot{r}^{6}+V(r)=0\nonumber
\end{eqnarray}
where the exact forms of the functions $X(r)$, $Y(r)$, $Z(r)$ and $V(r)$ depend on $f_{in}(r)$ and $f_{out}(r)$.

Therefore, we can no longer use quadratic particle dynamics
analogue to characterize the solutions of $r(t)$, instead it is
far easier to use phase space analysis. Let us parameterize the
phase space as $(q=r,p=\dot{r})$, and the Hamiltonian be

\be
\label{hamiltonGB} H=X(q)p^{2}+Y(q)p^{4}+Z(q)p^{6}+V(q)\;,
\ee
then the Hamilton's equations are

\bea
\label{hameq1} \dot{q}&=&2X(q)p+4Y(q)p^{3}+6Z(q)p^{5}\;,
\\
- \dot{p} &=& \frac{\partial X}{\partial q}p^{2}+\frac{\partial
Y}{\partial q}p^{4}+\frac{\partial Z}{\partial
q}p^{6}+\frac{\partial }{\partial q}V \;. \label{hameq2}
\eea
Given $f_{in,out}(r)$, we will then numerically solve these equations to characterize the solutions.

To find the explicit forms of $X$, $Y$, $Z$ and $V$, we assume the
spherical symmetry so that bulk and wall metrics again take the
forms of \eq{metricb} and \eq{metricw}, respectively. However, the
inside and outside harmonic functions solved the field equations
of Gauss-Bonnet gravity are the Boulware-Deser-Schwarzschild-de
Sitter ones \cite{Boul:1985}, that is,

\be \label{finGB}
f_{in}(r)=1+\frac{r^{2}}{4\alpha}(1-\sqrt{1+\frac{4\alpha\lambda}{3}})
\ee
where $\lambda$ is the cosmological constant inside the
bubble; and

\be \label{foutGB}
f_{out}(r)=1+\frac{r^{2}}{4\alpha}(1-\sqrt{1+\frac{4\alpha\Lambda}{3}+\frac{8\alpha\mu}{r^{4}}})
\ee
where $\Lambda$ is the
cosmological constant outside the bubble, and $\mu$ is the mass of
the bubble. If $\lambda$ is zero, this is the true vacuum bubble. On the other
hand, the false vacuum bubble has $\Lambda=0$.

Given \eq{finGB} and \eq{foutGB}, the functions $X$, $Y$, $Z$ and
$V$ appearing in \eq{hamiltonGB} for Hamilton formalism are \bea
X(r)&=&-32\alpha^{2}\Lambda^{2}+384\alpha
k^{2}+64\alpha^{2}\Lambda\lambda-32\alpha^{2}\lambda^{2}
-\frac{1152\alpha^{2\mu^{2}}}{r^{8}}-\frac{288\alpha\mu^{2}}{r^{6}}
-\frac{384\alpha^{2}\Lambda\mu}{r^{4}}+\frac{384\alpha^{2}\lambda\mu}{r^{4}}+\frac{768\alpha^{2}k^{2}}{r^{2}}\nonumber
\\
&&-\frac{96\alpha\Lambda\mu}{r^{2}}-\frac{48\alpha
k^{2}\mu}{r^{2}}+\frac{96\alpha\lambda\mu}{r^{2}}-8\alpha\Lambda^{2}r^{2}+36k^{2}r^{2}-8\alpha\Lambda
k^{2}r^{2}+16\alpha\Lambda\lambda r^{2}-8\alpha\lambda
k^{2}r^{2}-8\alpha\lambda^{2}r^{2}
\nonumber
\\
&&+\Bigl(-2r^{2}\Lambda-\frac{8}{3}r^{2}\alpha\Lambda^{2}+2r^{2}\lambda
+\frac{8}{3}r^{2}\alpha\lambda\Lambda-\frac{12\mu}{r^{2}}-\frac{32\alpha\Lambda\mu}{r^{2}}
-\frac{96\alpha\mu^{2}}{r^{6}}+\frac{16\alpha\lambda\mu}{r^{2}}\Bigr)\sqrt{1+\frac{4\alpha\Lambda}{3}
+\frac{8\alpha\mu}{r^{4}}}\nonumber
\\
&&+\Bigl(\frac{12\mu}{r^{2}}+\frac{16\alpha\lambda\mu}{r^{2}}+2\Lambda
r^{2}-2\lambda r^{2}+\frac{8}{3}\alpha\Lambda\lambda
r^{2}-\frac{8}{3}\alpha\lambda^{2}r^{2}\Bigr)\sqrt{1+\frac{4\alpha\lambda}{3}}\;,
\label{X1} \eea \be Y(r)=-16\alpha^{2}\Lambda^{2}+192\alpha
k^{2}+32\alpha^{2}\Lambda\lambda-16\alpha^{2}\lambda^{2}
+\frac{768\alpha^{2}k^{2}}{r^{2}}-\frac{192\alpha^{2}\Lambda\mu}{r^{4}}
+\frac{192\alpha^{2}\lambda\mu}{r^{4}}-\frac{576\alpha^{2}\mu^{2}}{r^{8}}\;,
\ee \be Z(r)=\frac{256\alpha^{2}k^{2}}{r^{2}}\;, \ee \bea
V(r)&=&-16\alpha^{2}\Lambda^{2}+192\alpha
k^{2}-\frac{3\mu}{2\alpha}+32\alpha^{2}\Lambda\lambda-16\alpha^{2}\lambda^{2}
-16\Lambda\mu-\frac{8\alpha\Lambda^{2}\mu}{3}
-12k^{2}\mu+12\lambda\mu-\frac{576\alpha^{2}\mu^{2}}{r^{8}}\nonumber
\\
&&-\frac{32\alpha\mu^{3}}{r^{8}}-\frac{288\alpha\mu^{2}}{r^{6}}-\frac{192\alpha^{2}\Lambda\mu}{r^{4}}
+\frac{192\alpha^{2}\lambda\mu}{r^{4}}-\frac{48\mu^{2}}{r^{4}}
-\frac{16\alpha\Lambda\mu^{2}}{r^{4}}+\frac{256\alpha^{2}k^{2}}{r^{2}}-\frac{96\alpha\Lambda\mu}{r^{2}}
-\frac{48\alpha k^{2}\mu}{r^{2}}\nonumber
\\
&&+\frac{96\alpha\lambda\mu}{r^{2}}-8\alpha\Lambda^{2}r^{2}+36k^{2}r^{2}-8\alpha\Lambda
k^{2}r^{2}+16\alpha\Lambda\lambda r^{2}-8\alpha k^{2}\lambda
r^{2}-8\alpha \lambda^{2} r^{2}-\frac{r^{4}}{8\alpha^{2}}\nonumber
\\
&&-\frac{\Lambda
r^{4}}{\alpha}-\frac{4}{3}\Lambda^{2}r^{4}-\frac{4}{27}\alpha\Lambda^{3}r^{4}
+\frac{k^{2}r^{4}}{\alpha}-2\Lambda
k^{2}r^{4}-k^{4}r^{4}-\frac{\lambda r^{4}}{4\alpha}+2\Lambda\lambda
r^{4}-2k^{2}\lambda
r^{4}-\frac{4}{3}\lambda^{2}r^{4}-\frac{4}{27}\alpha\lambda^{3}r^{4}
\nonumber
\\
&&+\Bigl(-\frac{3\mu}{\alpha}-8\alpha\Lambda-4k^{2}\mu+4\lambda\mu-\frac{96\alpha\mu^{2}}{r^{6}}
-\frac{24\mu^{2}}{r^{4}}-\frac{12\mu}{r^{2}}-\frac{32\alpha\Lambda\mu}{r^{2}}+\frac{16\alpha\lambda\mu}{r^{2}}-2\Lambda
r^{2}-\frac{8}{3}\alpha\Lambda^{2}r^{2}\nonumber
\\
&&+2\lambda r^{2}+\frac{8}{3}\alpha\Lambda\lambda
r^{2}-\frac{\Lambda
r^{4}}{2\alpha}-\frac{2}{3}\Lambda^{2}r^{4}-\frac{k^{2}r^{4}}{2\alpha}-\frac{2}{3}\Lambda
k^{2}r^{4}+\frac{\lambda r^{4}}{2\alpha}+\frac{2}{3}\Lambda\lambda
r^{4}\Bigr)\sqrt{1+\frac{4\alpha\Lambda}{3}+\frac{8\alpha\mu}{r^{4}}}
\nonumber
\\
&&+\Bigl(\frac{3\mu}{\alpha}+4\lambda\mu+\frac{12\mu}{r^{2}}+\frac{16\alpha\lambda\mu}{r^{2}}+2\Lambda
r^{2}-2\lambda r^{2}+\frac{8}{3}\alpha\Lambda\lambda
r^{2}-\frac{8}{3}\alpha\lambda^{2}r^{2}\nonumber
\\
&&+\frac{\Lambda
r^{4}}{2\alpha}-\frac{k^{2}r^{4}}{2\alpha}-\frac{\lambda
r^{4}}{2\alpha}+\frac{2}{3}\Lambda\lambda
r^{4}-\frac{2}{3}k^{2}\lambda
r^{4}-\frac{2}{3}\lambda^{2}r^{4}\Bigr)\sqrt{1+\frac{4\alpha\lambda}{3}}
\nonumber
\\
&&+\Bigl(\frac{\mu}{\alpha}+\frac{4}{3}\lambda\mu+\frac{r^{4}}{8\alpha^{2}}+\frac{\Lambda
r^{4}}{6\alpha}+\frac{\lambda
r^{4}}{6\alpha}+\frac{2}{9}\Lambda\lambda
r^{4}\Bigr)\sqrt{1+\frac{4\alpha\lambda}{3}}\sqrt{1+\frac{4\alpha\Lambda}{3}
+\frac{8\alpha\mu}{r^{4}}}\;, \label{V1} \eea where the re-scaled
tension parameter \be k:=\kappa^2 \sigma. \ee These forms are so
complicated so that we can only rely on the numerical analysis on
the phase space to characterize various kind of solutions by tuning
the parameters and initial conditions.

\subsubsection{Breathing true vacuum bubble}

\begin{figure}
\epsfxsize=300pt \centerline{\epsfbox{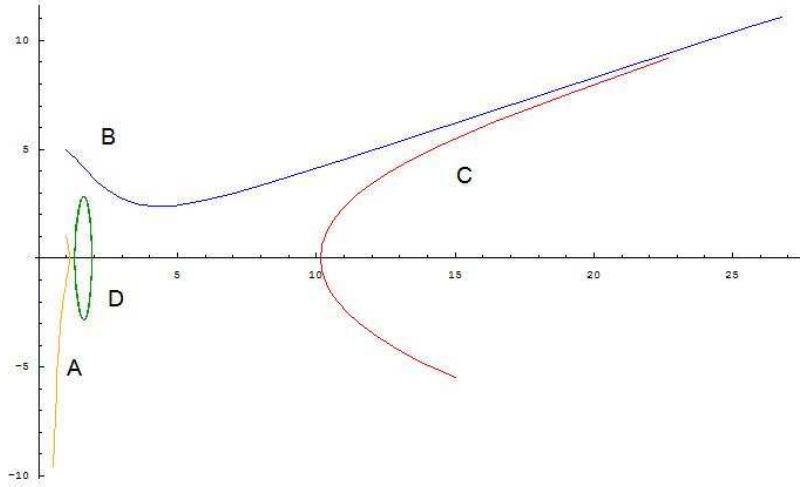}} \caption{True vacuum
bubble solutions on phase space for 5D Gauss-Bonnet gravity.  A:
bounded solution. B: monotonic solution. C: bounce solution. D:
breathing solution. }\label{fig:3}
\end{figure}

We first consider the dynamics of true vacuum bubble, which could
be relevant to the global picture of string landscape. To
implement numerical analysis of the Hamilton equations \eq{hameq1}
and \eq{hameq2}, we set  $\Lambda=1$, $\lambda=0$, $\mu=1$ and
$\alpha=0.01$ in \eq{X1}-\eq{V1}. For generic values of $k$, we
will as usual have bounded, bounce and monotonic solutions, but
for some regime of $k$ we have the breathing bubble solution,
i.e., the size of the bubble is pulsating/breathing so that the
phase space trajectory is periodic. The general numerical
solutions on phase space are illustrated in Figure \ref{fig:3}.

\begin{figure}
\epsfxsize=300pt \centerline{\epsfbox{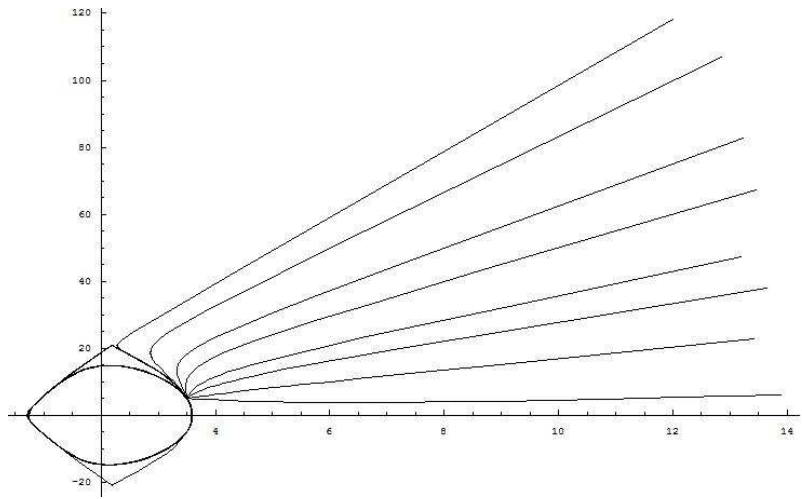}} \caption{Tuning the
re-scaled tension parameter $k$ for breathing bubble. We fix
$\Lambda=1$, $\mu=1$, $\alpha=0.01$, and the initial condition
$p=5$ and $q=3.5$ at which all the lines intersect in the phase
space diagram. The solution corresponding to the almost horizontal
line has $k=1$, and then the value of $k$ decreases for the lines
arranged counter-clockwise until $k=0.0253$ below that there
exists the  breathing solution, i.e., the elliptical shape
ones.}\label{fig:4}
\end{figure}

To examine more closely on the breathing solution, we tune the
re-scaled tension parameter $k$ for a given some initial
condition, and find that there is a critical value of $k$, lower
than that there will exist the breathing solution. The dependence
on the initial condition of bubble's phase space trajectory  is
shown in Figure \ref{fig:4}.

\subsubsection{No breathing false vacuum bubble}

\begin{figure} \epsfxsize=400pt \centerline{\epsfbox{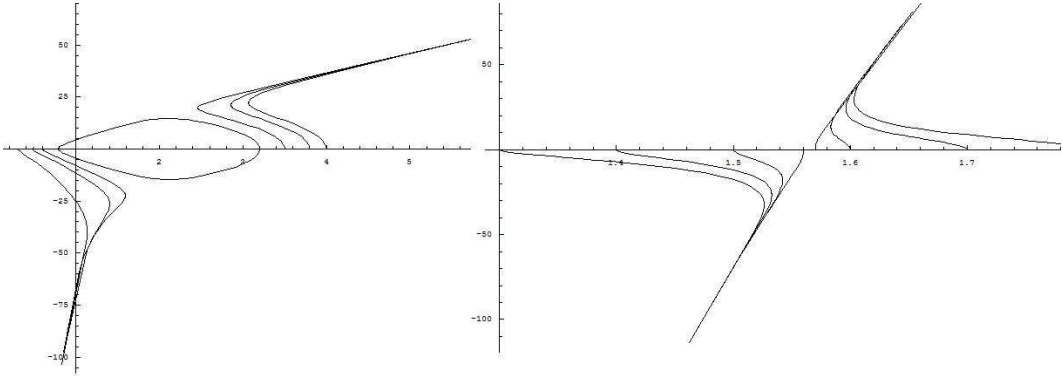}}
\caption{Phase space diagram for tuning the initial conditions for
the true (Left) and the false (Right) vacuum bubbles. On the left,
we set $\alpha=0.01$, $k=0.027$, $\mu=1$, $\Lambda=1$ and
$\lambda=0$. On the right, we set $\alpha=0.01$, $k=0.001$,
$\mu=1$, $\Lambda=0$ and $\lambda=1$. We see there is a breathing
solution for the true vacuum bubble, but not for the false one.
}\label{fig:5}
\end{figure}

We can also consider the dynamics of the false vacuum by choosing
$\Lambda=0$ and $\lambda= 1$ for the harmonic functions \eq{finGB}
and \eq{foutGB}. Again, solving the Hamilton equations
\eq{hameq1}-\eq{hameq2} with the set of functions in
\eq{X1}-\eq{V1} in terms of the new harmonic functions, we find
that there is no breathing solution. For illustration, in Figure
\ref{fig:5} we juxtapose the phase space diagrams by tuning the
initial conditions for the true and false vacuum bubbles.

We can also consider the cases with both $\Lambda$ and $\lambda$
non-zero, then it is more close to true vacuum bubble if
$\Lambda>\lambda$, and more close to false vacuum bubble if
$\Lambda<\lambda$. From our numerical studies with quite extensive
ranges for both parameters and initial conditions, we find that
there exist breathing bubble solutions for the former case, but
not for the latter. We may then tend to state that there are
breathing vacuum bubble solutions in 5D Gauss-Bonnet gravity if
the cosmological constant inside the bubble is smaller than the
one outside, but not vice versa.

\section{Conclusion and the implications}

The higher derivative terms in Einstein-Hilbert action can be
ignored when the length scale considered is much smaller than the
curvature radius. Thus there is no reason to dwell on the Einstein
gravity when we consider the global picture of the universe. In
this case it will be important to exam the effect of the higher
derivative corrections. One new feature we found in this paper is
the breathing bubble solution in 5D Gauss-Bonnet gravity. In
particular, we don't need the exotic matter for the bubble wall
and this kind of solutions exist for true vacuum bubble with zero
or positive vacuum energy. One possible implication is on the
measure problem of eternal inflation. In order to do the sensible
statistics on bubble distribution, we need to cut-off the infinite
space-time and infinite number of bubbles properly. One way of
doing this is to count the bubbles smaller than a fixed scale
factor \cite{Garriga:2005}. However usually people assume the
bubble nucleating at vacuum with Hubble constant $H$ will have
co-moving radius $H^{-1}$ asymptotically. This is not the case if
there are breathing bubbles whose co-moving sizes are vary with
time and depend on critical bubble size. Including these bubbles
will change the evolution equation for the fraction of volume
occupied by a particular bubble. One may also ask what happens if
a growing bubble nucleates inside a breathing bubble. The bubble
walls can collide and generate the gravitational wave. This can
have the observational consequences if our universe goes through
such kind of bubbles in the past evolution. We leave these
problems for the future studies.

\section*{Acknowledgements}
  This work is supported by Taiwan's NSC grant 097-2811-M-003-012 and 97-2112-M-003-003-MY3.


\appendix

\section{Difficulty for thin-wall vacuum bubbles in higher derivative gravity }
One difficulty of studying bubble dynamics in arbitrary higher
derivative gravity is due to the appearance of higher order
singularity when assuming the thin wall approximation. In
\cite{Sasaki:2007}, they try to avoid this difficulty by assuming
the discontinuity of extrinsic curvature only appears in higher
order derivative. However we don't have solutions satisfing this
criterion yet. In this appendix, we will derive the junction
conditions modified by higher order corrections and argue that in
general they are not the consistent sets of equations.

A Lagrangian including higher order corrections derived from high
energy theory like string theory may have the form
\begin{eqnarray}
\textit{L}=R-2\Lambda+\alpha R^2+\beta R_{\mu\nu}R^{\mu\nu}+\gamma
R_{\alpha\beta\gamma\delta}R^{\alpha\beta\gamma\delta}\nonumber
\end{eqnarray}
A special case for second order gravity is
$\alpha=\gamma=1,\beta=-4$ which is the Einstein-Gauss-Bonnet
gravity. In four dimensions these terms are topological invariant
which does not contribute to the equation of motion. For
simplicity, we consider $\emph{L}^{(2)}=R-2\Lambda+\alpha R^2$
first and show that thin wall approximation is not reliable. The
field equation from variation of $\sqrt{g}\emph{L}$
is \cite{Sasaki:2007}
\begin{eqnarray}
\sigma_{\mu\nu}=(1+2\alpha R)G_{\mu\nu}+\frac{1}{2}\alpha
R^{2}g_{\mu\nu}-2\alpha D_{\mu}D_{\nu}R+2\alpha g_{\mu\nu}\Box
R^{2}\nonumber
\end{eqnarray}
\begin{eqnarray}
\sigma_{ij}=\Lambda h_{ij}+G_{ij}+2\alpha
RG_{ij}+\alpha\frac{1}{2}R^{2}h_{ij}+2\alpha[{-\overline{D}_{i}}\overline{D}_{j}R
+K_{ij}\partial_{y}R+h_{ij}(\partial_{yy}R+\overline{\Box}R-K\partial_{y}R)]\nonumber
\end{eqnarray}
In order to extract the discontinuity from equation of motion, we
assume the metric is continuous at the wall, but it has a kink.
Its first derivative has a step function discontinuity and its
second derivative has a delta function term \cite{8}
\begin{eqnarray}
h_{ij}(y)=h_{ij}^{-} \theta(-y)+h_{ij}^{+} \theta(y)\nonumber
\end{eqnarray}
\begin{eqnarray}
\frac{\partial h_{ij}(y)}{\partial y}=\frac{\partial
h_{ij}^{-}(y)}{\partial y}\theta(-y)+\frac{\partial
h_{ij}^{+}(y)}{\partial y}\theta(y)\nonumber
\end{eqnarray}
\begin{eqnarray}
\frac{\partial^{2} h_{ij}(y)}{\partial^{2} y}=\frac{\partial^{2}
h_{ij}^{-}(y)}{\partial^{2} y}\theta(-y)+\frac{\partial
h_{ij}^{+}(y)}{\partial y}\theta(y)+(\frac{\partial
h_{ij}^{-}(y)}{\partial y}+\frac{\partial h_{ij}^{+}(y)}{\partial
y})\delta(y)\nonumber
\end{eqnarray}
In Einstein's gravity, substitute (24) into
$G_{ij}=\kappa^{2}T_{ij}$ and match the delta function term, so
that we can get the Israel junction condition. In order to get the
junction condition for the $R^{2}$ correction gravity, we need to
work out following quantity
\begin{eqnarray}
~&&~RG_{ij}
=2\partial_{y}K\partial_{y}(K_{ij}-Kh_{ij})+2\partial_{y}K\Bigr(2K_{i}^{l}K_{lj}+\partial_{y}K_{lj}-KK_{ij}
+\frac{1}{2}h_{ij}(TrK^{2}+K^{2})+\bar{G}_{ij}\Bigl)\nonumber\\&&~
+\partial_{y}(K_{ij}-Kh_{ij})(-TrK^{2}-K^{2}+\bar{R})
\end{eqnarray}
\begin{eqnarray}
R^{2}=[2\partial_{y}K-TrK^{2}-K^{2}+\bar{R}]^{2}
\end{eqnarray}
\begin{eqnarray}
\partial_{y}RK_{ij}=[2\partial_{yy}K-\partial_{y}TrK^{2}-\partial_{y}K^{2}+\partial_{y}(\bar{R})]K_{ij}
\end{eqnarray}
\begin{eqnarray}
\partial_{yy}R=2\partial_{yyy}K-\partial_{yy}TrK^{2}-\partial_{yy}K^{2}+\partial_{yy}(\bar{R})
\end{eqnarray}
\begin{eqnarray}
K\partial_{y}R=K(2\partial_{yy}K-\partial_{y}TrK^{2}-\partial_{y}K^{2}+\partial_{y}(\bar{R})
\end{eqnarray}
We consider that induced metric $h_{ij}$ is continuous across the
domain wall, but discontinuous across the wall. However terms like
$R^{2}$ would give $\delta_(y)^{2}$ and $\partial_{y}R $ would
give $\partial_{y}\delta(y)$ and $\partial _{yy}R$ would give
$\partial_{yy}\delta(y)$. Those more singular terms are not well
defined. One may require higher continuity of the induced metric
$h_{ij}$, but this would make the equation of the domain wall
trivial. We cannot assume this condition in our case. We propose
to include more singular source terms on the wall to see if the
equation of motion can be solved consistently. Introducing source
terms
\begin{eqnarray}
T^{\mu \nu}=S^{\mu \nu}(x^{i})\delta(y)+U^{\mu
\nu}(x^{i})\delta(y)^{2}+V^{\mu
\nu}(x^{i})\partial_{y}\delta(y)+W^{\mu
\nu}(x^{i})\partial_{yy}\delta(y)\nonumber
\end{eqnarray}
Due to the conservation law$\nabla _{\nu} T^{\mu \nu}=0$
\begin{eqnarray}
\nabla _{\nu} T^{i \nu}=(\nabla _{j}S^{ij}+2K_{j}^{i}S^{jy}+TrKS^{iy})\delta(y)+S^{iy}\partial_{y}\delta(y)\nonumber\\
+(\nabla _{j}U^{ij}+2K_{j}^{i}U^{jy}+TrKU^{iy})\delta(y)^{2}+U^{iy}2\delta(y)\partial_{y}\delta(y)\nonumber\\
+(\nabla _{j}V^{ij}+2K_{j}^{i}V^{jy}+TrKV^{iy})\partial_{y}\delta(y)+V^{iy}\partial_{yy}\delta(y)\nonumber\\
+(\nabla
_{j}W^{ij}+2K_{j}^{i}W^{jy}+TrKW^{iy})\partial_{yy}\delta(y)+W^{iy}\partial_{yyy}\delta(y)\nonumber\\=0\nonumber
\end{eqnarray}
We should have the following constraints.
\begin{eqnarray}
\nabla _{j}S^{ij}+2K_{j}^{i}S^{jy}+TrKS^{iy}=0\nonumber\\
\nabla _{j}U^{ij}+2K_{j}^{i}U^{jy}+TrKU^{iy}=0\nonumber\\
S^{iy}+\nabla _{j}V^{ij}+2K_{j}^{i}V^{jy}+TrKV^{iy}=0\nonumber\\
U^{iy}=0\nonumber\\
\nabla _{j}W^{ij}+2K_{j}^{i}W^{jy}+TrKW^{iy}+V^{iy}=0\nonumber\\
W^{iy}=0\nonumber
\end{eqnarray}
Matching the left hand side and right hand side of equation for
the first and second derivatives of the delta function and the
square of the delta function respectively. It gives four
equations, where one being proportional to $\delta(y)$ is rather
cumbersome, so we neglect it, but it will not affect our result
\begin{eqnarray}
\partial_{yy}\delta(y)\longrightarrow -2\alpha h^{ij+}(\partial h_{ij}^{+}-\partial h_{ij}^{-}) h_{ij}=W_{ij}(x^{i})\nonumber
\end{eqnarray}
\begin{eqnarray}
~&&~\delta(y)^{2}\longrightarrow
\frac{\alpha}{2}h^{+kl}(\partial_{y}
h_{kl}^{+}-\partial_{y}h_{kl}^{-})(\partial_{y}
h_{ij}^{+}-\partial_{y}h_{ij}^{-})+\frac{\alpha}{2}[h^{+kl}(\partial_{y}
h^{+}_{kl}-\partial_{y} h^{-}_{kl})h^{+mn}(\partial_{y}
h^{+}_{mn}-\partial_{y} h^{-}_{mn})]h_{ij}\nonumber\\&&~ +
\frac{\alpha}{4}[h^{kl+}(\partial_{y} h^{+}_{kl}-\partial_{y}
h_{kl}^{-})h^{+mn}(\partial_{y} h^{+}_{mn}-\partial_{y}
h^{-}_{mn})(\partial_{y} h^{+}_{ij}+\partial_{y}
h^{-}_{ij})]+\frac{\alpha}{2}[h^{+ij}(\partial_{y}
h_{ij}^{+}-\partial_{y} h_{ij}^{-})]^{2}h_{ij}=U_{ij}\nonumber
\end{eqnarray}
\begin{eqnarray}
~&&~\partial_{y}\delta(y)\longrightarrow  \frac{\alpha}{2}
[h^{+kl}\partial_{y}h_{ij}^{+}\partial_{y}
h_{kl}^{+}+h^{+kl}\partial_{y}h_{ij}^{-}\partial_{y}+h^{+ij}(h^{+kl}\partial_{y}h_{ij}^{+}\partial_{y}
h_{kl}^{+}+h^{+kl}\partial_{y}h_{ij}^{-}\partial_{y}+
h_{kl}^{-})h_{ij}\nonumber\\&&~-2(\partial_{y} h^{+ij}\partial_{y}
h_{ij}^{+}+\partial_{y} h^{+ij}\partial_{y}
h_{ij}^{-}+h^{+ij}\partial_{y}^{2} h_{ij}^{+}+h^{+ij}\partial_{y}
h_{ij}^{-}+\partial_{y} h^{+ij}(\partial_{y}
h_{ij}^{+}-\partial_{y}
h^{-}_{ij})\nonumber\\&&~+h^{+ij}(\partial_{y}
h_{ij}^{+}-\partial_{y} h_{ij}^{-})+\partial_{y}
h^{+ij}\partial_{y} h^{+}_{ij}+h^{+ij}\partial_{y}^{2}
h_{ij}^{+}-\partial_{y} h^{-ij}\partial_{y}
h^{-}_{ij}-h^{-ij}\partial_{y} h^{-}_{ij})h_{ij}
\nonumber\\&&~+(h^{+ik}h^{+jl}(\partial_{y} h_{ij}^{+}\partial_{y}
h_{kl}^{+}-\partial_{y} h^{-}_{ij}\partial_{y}
h_{kl}^{-})+h^{+ij}h^{+kl}(\partial_{y} h_{ij}^{+}\partial_{y}
h_{kl}^{+}-\partial_{y} h^{-}_{ij}\partial_{y}
h_{kl}^{-}))h_{ij}]=V_{ij}\nonumber
\end{eqnarray}
We can see that these three equations are very different and
cannot be solved simultaneously. It shows the breakdown of the
thin wall approximation. Instead, we should consider finite
thickness of the domain wall to avoid severe singular behavior.


 \end{document}